\documentstyle[aps,prl,psfig]{revtex}
\newcommand{\be}{\begin{equation}}
\newcommand{\ee}{\end{equation}}

\begin{document}
\title{\bf Money and Goldstone modes}
\author{Per Bak$^{1,2}$, Simon F. N\o rrelykke$^2$, and Martin Shubik$^3$}
\address{$^1$Imperial College, London WS7 2BZ, UK}
\address{$^2$Niels Bohr Institute, Blegdamsvej 17, 2100 Copenhagen, Denmark.}
\address{$^3$Cowles Foundation for Research in Economics, 
	Yale University, PO Box 208281, New Haven, CT 06520, USA}
\date{\today}
\maketitle

\begin{abstract}
Why is  ``worthless'' fiat money generally accepted as payment for goods
and services? In equilibrium theory, the value of money is generally not
determined: the number of equations is one less than the number of unknowns,
so only relative prices are determined.
In the language of mathematics, the equations are ``homogeneous of order one''.
Using the language of physics, this represents a continuous ``Goldstone''
symmetry. However, the continuous symmetry is often broken by the dynamics
of the system, thus fixing the value of the otherwise undetermined variable.
In economics, the value of money is a strategic variable which each
agent must determine at each transaction by estimating the effect of
future interactions with other agents. 
This idea is illustrated by a simple network model of 
monopolistic vendors and buyers, with bounded rationality. 
We submit that dynamical, spontaneous symmetry breaking is the fundamental
principle for fixing the value of money. Perhaps the continuous symmetry 
representing the lack of restoring force is also the fundamental reason for 
large fluctuations in stock markets.
\end{abstract}


%%%%%%%%%%%%%%%%%%%%%%%%%%%%%%%%%%%%%%%%%%%%%%%%%%%%%%%%%%%%%%%%
	
\section{Introduction}

In classical equilibrium theory in economics~\cite{Debreu}, agents submit 
their demand-versus-price functions to a ``central agent'' who then 
determines the 
relative prices of goods and their allocation to individual agents.
The absolute prices are not fixed, so the process does not determine the 
value of money, which merely enters as a fictitious quantity that
facilitates the calculation of equilibrium. Thus, traditional equilibrium 
theory does not offer a fundamental explanation 
of money, perhaps the most essential quantity in a modern economy. 

Money is essentially a dynamical phenomenon, since its role is intimately 
related to the temporal sequence of events. Suppose an 
agent has apples and wants oranges. He might have to sell his apples to 
another agent before he buys oranges from a third agent: hence money is needed 
for the transaction, supplying liquidity. It stores value between transactions.
If everybody could get together at a common market-place, there would be no
need for money.

A ``search-theoretic'' approach to monetary economics has recently 
been proposed,~\cite{KW,TW}, in which agents may be either money 
traders, producers, or commodity traders. 
The agents randomly interact with each other, and
they decide whether or not to trade based on ``rational expectations''
about the value of a transaction. After a transaction the agent 
changes into one of the two other types of agents. 
This theory has a steady state where money circulates. 
As in other equilibrium theories, 
this theory does not describe a dynamics
leading to the steady state, of sufficient detail, say for one to simulate it
if one would so desire.  
The value of money comes about by: (1) pre-defining
the ratio of agents which are money traders; and (2) requiring each 
money-trader to spend all his money at each transaction. 
Actually, Trejos and Wright~\cite{TW} 
found their monetary equilibrium to be unstable with respect to a 
small perturbation of the expected value of money, 
causing the system to slide
to a stable fixed point where money has no value! 

Within our picture~\cite{1999} there is no restoring force for the value
of money---which could in principle be anything, with no consequences for
the actual physical transactions that takes place. The value of 
money represents a ``continuous symmetry''. If, at some point, the value
of money was globally redefined by a certain factor, this would have no
consequences whatsoever. Thus, in order to arrive at a specific value
of money, the continuous symmetry must be broken. 
 
The principle of continuous symmetries, and, particularly, the 
spontaneous breaking of those symmetries, is hugely important in  
physics. Indeed, it is the basic principle behind our understanding of the 
properties of fundamental particles. It is also responsible for magnetic 
order and superconductivity. 
As an example, consider, for instance, a lattice of interacting atoms 
forming a crystal. The crystal's physical properties, including its energy,  
are not affected by a uniform translation $X$ of all atoms.  The 
position of the crystal represents a continuous symmetry. It could in principle 
be anywhere. Nevertheless, starting with an arbitrary position of the 
individual atoms, the dynamics of the atoms interacting with their 
neighbors fix the actual position of the entire crystal! We say 
that the continuous symmetry is ``dynamically broken''. Note that if we shift 
the crystal there is no restoring force. Also, if the system is subjected 
to random noise, the lack of restoring force, expressed by the underlying  
continuous symmetry, leads to large positional fluctuations. The 
atoms exhibit ``bounded rationality'': they simply adjust to the positions of 
their neighbors, wherever they might me. In order to calculate their  
dynamics one does not need any 
information of the ultimate equilibrium (if one exists) of the entire system. 
 
In economics, the agents represent the atoms of the economy. The value that 
an agent assigns to money corresponds to the position of an atom.  
The neighbors represent the subsystem of agents with which an individual agent 
interact, in one way or another. The local actions of the agents determine 
the value of money for each agent at each instant of time.  
The average value of  
money, corresponding to the ``center of mass'' for the atomic chain, is fixed  
by the collective dynamics of all the agents. 
 
Figure 1 illustrates the concept of continuous symmetries in the two 
cases. The physical properties of the chain of atoms are invariant under 
a uniform shift of all atoms. The economy is invariant under a uniform 
shift of all prices. The actual positions, including the 
``center of mass'' in both cases are fixed by the dynamics.

\begin{figure}
\centerline{\psfig{figure=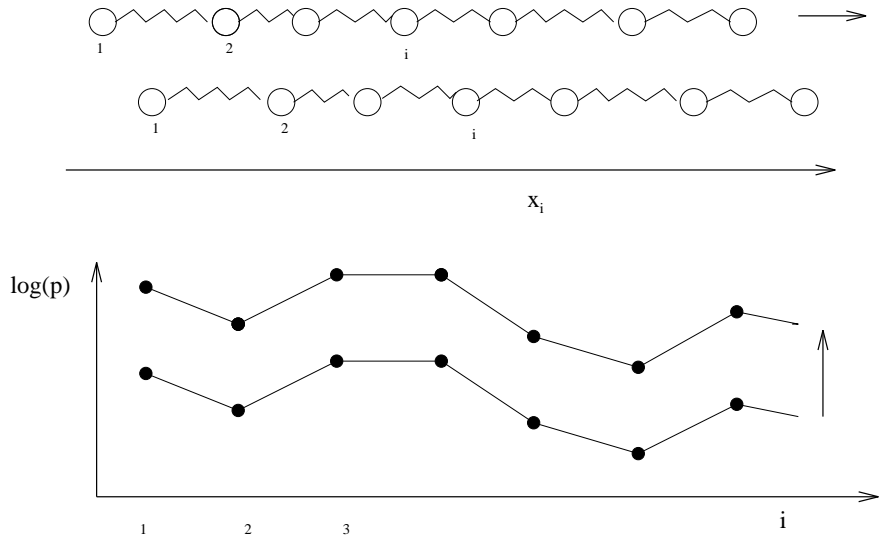,angle=270,height=7cm}}
\vspace{3mm}
\caption{Continuous symmetries for chain of atom (top) and 
	prices (bottom). The properties of both systems are invariant under 
	a uniform shift as indicated by the arrow.}  
\end{figure}

It is hard to envision a picture of an economy  
which does not derive its properties starting from a reasonably realistic and 
``simulatable'' model of the behavior of the individual agents.  
In the final analysis,  
even equilibrium theory must have a dynamical underpinning 
if, in any sense, it aims at modeling a dynamical world. 
 
It is not so important for our considerations how the agents actually behave, 
i.e., to what extent their behavior might be considered rational, as long 
as they somehow adjust their behavior in response to their local environment.  
To be specific,  
we study a network of vendors and buyers, each of whom have a simple  
local optimization strategy. Whenever a transaction is considered, 
the agent must decide the value of the goods and services in question,  
or, equivalently, the value of money relatively to that of the goods  
and services he intends to buy or sell. He will associate that value 
to his money that he believes will maximize his utility. Thus, the  
value of money is a ``strategic variable,'' that the agent in principle 
is free to choose as he pleases. However, if he makes a poor choice he will 
loose utility. In any case, the dynamics of his action, together with those 
of other agents, will break the continuous symmetry. 
 
For simplicity, we assume that agents are rather myopic: they have short  
memories, and they take into account only the properties of their  
``neighbors,'' i.e., the agents with which they interact directly. They 
have no idea about what happens elsewhere in the economy. 
 
Despite the bounded rationality of these agents,  
the economy self-organizes into an equilibrium state. 
Since we define the dynamics explicitly, we are, however, able to treat 
the nature of this relaxation to the equilibrium state, as well as  
the response of the system to perturbations, and to noise-induced  
fluctuations around the equilibrium.  
These phenomena are intimately related to the 
dynamics of the system, and cannot be discussed within  
any theory concerned only with the equilibrium situation.   
The value of money is fixed by a ``bootstrap'' process: agents  
benefit from assigning a specific value to money,  
despite this value's global indeterminacy.   
                          
\section{The Model} 
In our model, we consider $N$ agents, $n = 1,2,\ldots ,N$, placed on a 
one-dimensional lattice with periodic boundary conditions. This geometry 
is chosen in order to have a simple and specific way of defining who is  
interacting with whom. The geometry is not important for our general  
conclusions concerning the principles behind the fixation of prices. 
 
We assume that agents cannot consume their own output, so in order to consume 
they have to trade, and in order to trade they need to produce. 
Each agent  produces a quantity $q_{n}$, of one good, which is sold 
at a unit price $p_n$, to his left neighbor $n-1$.  
He next buys and consumes one good from his neighbor to the right, who  then  
buys the good of {\em his\/} right neighbor, etc., until all agents have made 
two transactions. This process is repeated indefinitely, say, once per day. 
 
Our model is a simple extension of Jevons'~\cite{Jevons} 
example of a three  agent, 
three commodity economy with the failure of the 
double coincidence of  wants, i.e., 
when only one member of a trading pair wants a good owned by the other. 
A way out of the paradox of no trade
where there is gain to be obtained by all 
is to utilize a money desired by and held by all.  
Originally this was gold, 
but here we show that the system dynamics can attach value to 
``worthless'' paper money. 
 
For simplicity, all agents are given utility 
functions of the same form 
 
\be 
 u_n = -c(q_{n}) + d(q_{n+1}) +  
	     I_{n}\! \cdot \! (p_{n}q_{n}- p_{n+1}q_{n+1}) \enspace. 
  \label{eq:utility} 
\ee 
 
The first term $-c$, represents the agent's cost, or displeasure, 
associated with producing $q_{n}$ units of the good he produces.  
The displeasure is an increasing  
function of $q$, and $c$ is convex, say because the agent gets  
tired. The second term $d$, is his utility of the good he  
can obtain from his neighbor. 
Its marginal utility is decreasing with $q$,  
so $d$ is concave.  
 
An explicit example  
is chosen for illustration and analysis, 
 
\begin{equation} 
  c(q_n) =  a q_n^\alpha  \mbox{\hspace{0.5em},\hspace{1.5em}}  
  d(q_{n+1}) = b q_{n+1}^\beta \enspace , 
\end{equation} 
where $a$ is $\frac{1}{2}$ and of dimension [utility per (unit of $q_n$)$^{\alpha}$], 
$b$ is 2 and of dimension [utility per (unit of $q_{n+1}$)$^{\beta}$], 
and $\alpha$ and $\beta$ are chosen as 2 and $\frac{1}{2}$. The 
specific values of $a$, $b$, $\alpha$, and $\beta$ are 
not important for the general results, as long as $c$ remains  
convex and $d$ concave. 
 
The last term represents the change in utility associated with the gain 
or loss of money after the two trades. 
Notice that the dimension of $I_n$ 
is [utility per unit of currency], i.e., the physical interpretation 
is the {\em value of money.}  
 
Each agent has knowledge only  
(indirectly through the prices that were charged)  
about the utility functions of his two neighbors,  
as they appeared the day before. 
The agents are monopolistic, i.e., agent $n$ sets the 
price of his good, and agent $n-1$ then decide how much $q_{n}$, he  
will buy 
at that price. This amount is then produced and sold---there is no  
excess 
production. The goal of each agent is to maximize his utility,  
by adjusting $p_n$ and $q_{n+1}$, while 
maintaining a constant amount of money. Money has value only as  
liquidity.  
 
The agents aim to achieve a situation where the expenditures are  
balanced by the income: 
 
\be 
  p_{n}q_{n} -  p_{n+1}q_{n+1} = 0 \enspace . 
  \label{eq:constraint} 
\ee 
 
This condition is important for our considerations. The agent envisions 
an ongoing process, which will be repeated many times. In real life, 
he will eventually deal with very many agents. If he spends more 
money than he earns, he will eventually run out of money, or reach a ceiling 
for debt, causing loss of utility thereafter. If he uses less than he earns,  
he will be hoarding money, which gives no pleasure, but with a loss of  
utility. We assume that there is no preference for consumption now rather  
than later. Perhaps he will over-spend for a while, but eventually he 
will try to balance his budget. 
  
Note that when the value of money is fixed at a global value,  
$I_n$ = $I$, 
the agents optimize their utility by charging a price   
 
\be 
  p = 2^{\frac{1}{3}} \cdot I^{-1} 
 \label{eq:moneq1} 
\ee 
and selling an amount 
\be 
  q = 2^{-\frac{2}{3}} 
  \label{eq:moneq2} 
\ee 
at that price. This is the monopolistic equilibrium.  
In general, of course, 
with a different set of strategies there is no reason that the dynamics  
should lead to the monopolistic equilibrium,  
or to any stationary state for that sake. 
Note that the resulting quantities,  
Eq.~(\ref{eq:moneq2}), 
are independent of the value of money,  
which thus represents a continuous 
symmetry. There is nothing in the equations that fixes the 
value of money and the prices. Mathematically, the continuous symmetry 
expresses the fact that the equations for the quantities are 
``homogeneous of order one.'' The number of equations is one 
less than the number of unknowns, leaving the value of money 
undetermined. This is the continuous symmetry which 
eventually is broken by the dynamics. 
 
Agent $n$ tries to achieve his goal, 
of maximizing his utility and keeping constant his amount of money, 
by estimating 
the amount of goods $q_n$  his neighbor will order at a given price,  
and the price $p_{n+1}$ his other neighbor will charge at the 
subsequent transaction. 
Knowing that his 
neighbors are rational beings like himself, he is able to  
deduce the functional relationship between the price $p_n$, 
he demands and the amount of goods $q_n$, that will be  
ordered in response to this. 
Furthermore, he is able to make a guess at the  
size of $p_{n+1}$, based on the previous transaction with 
his right neighbor. This enables him to decide what the  
perceived value of money should be, and hence how much 
he should buy and what his price should be. 
This process is then continued indefinitely, 
at times $\tau = 1,2,3, \ldots $ 
 
This defines the game.  
The strategy we investigate contains the  
assumption made by each agent that the other agents do not  
change their valuation of money $I$,  
between their two daily transactions,  
and hence they maximize their utility accordingly. 
 
The process is initiated by choosing some initial values for 
the $I$s. They could, e.g., be related to some former gold  
standard. 
In fixing his price at his first transaction of day $\tau$, agent $n$  
exploits the knowledge he has 
of his neighbors' utility functions, i.e., he knows that the agent to  
the left will maximize his function with respect to $q_{n,\tau}$ 
 
\be 
  \frac{\partial u_{n-1,\tau}}{\partial q_{n,\tau}} =  0  
  \label{eq:maximize} 
  \enspace . 
\ee  
 
His knowledge of the functional relationship, between the amount 
of goods $q_{n-1,\tau}$, ordered by agent $n-1$ at time $\tau$ 
and the price $p_{n,\tau}$, set by agent $n$, allows agent $n$ 
to gauge the effect of his price policy. Lacking knowledge about 
the value of $I_{n-1,\tau}$, agent $n$ instead uses the value  
from the last transaction $I_{n-1,\tau-1}$.  
Substituting the expression for  
$q_{n,\tau}$, found from solving Eq.~(\ref{eq:maximize}), 
into Eq.~(\ref{eq:utility}) and 
maximizing with respect to $p_{n}$ and $q_{n+1}$ 
we find (for details see~\cite{1999}) 
 
\begin{equation} 
  I_{n,\tau+1} = \left( I_{n-1,\tau }^4 \, I_{n,\tau }^2 \, 
  I_{n+1,\tau }\right) ^{\frac{1}{7}} 
 \label{eq:In} 
\end{equation} 
which is a weighted geometric average of the value the two neighbors 
and agent $n$ 
prescribed to their money the previous day. Using this value of $I_{n}$,  
agent $n$ can fix his price $p_{n}$ and decide  
which quantity $q_{n+1}$, he should optimally buy. 
This simple equation completely specifies the dynamics of 
our model. The entire strategy can be reduced to an update scheme 
involving only the value of money---everything else follows from this. 
Thus, the value of money can be considered the basic strategic variable! 
 
Even though there is no utility in the possession of money,  
as explicitly expressed by Eq.~(\ref{eq:constraint}),  
the strategies and dynamics of the model nevertheless leads to a real 
utility, given by the third term in Eq.~(\ref{eq:utility}), being  
assigned to money. {\it The dynamics in this model is driven by the need of 
the agents to make estimates about future transactions.}  
In a sense, this models the real world where agents are forced to make  
plans about the future, based on knowledge about the past---and, in practise, 
only a very limited part of the past. 
 
In the steady state, where the homogeneity of the utility 
functions give $I_n = I_{n+1}$, we retrieve the monopolistic 
equilibrium equations, Eqs.~(\ref{eq:moneq1})~and~(\ref{eq:moneq2}).  
Of course, this is somehow accidental.  
Other more or less rational strategies will lead to 
different values of money.

\subsection{Stability analysis} 
 
Taking the logarithm and introducing $h_{n,\tau }= \ln (I_{n,\tau })$  
yields the linear equation: 
 
\begin{equation} 
  h_{n,\tau +1}= \mbox{\small $\frac{4}{7}$}h_{n-1,\tau } +   
  \mbox{\small$\frac{2}{7}$}h_{n,\tau} 
  +  \mbox{\small$\frac{1}{7}$}h_{n+1,\tau }\enspace .  
 \label{eq:log} 
\end{equation} 
This describes a Markov process. Clearly, any $h=constant$ solves 
this equation, in accordance with the fact that $h$ is a continuous 
symmetry. Thus, in order to get solutions, one must specify initial 
conditions.  
A simulation of 1000 agents with random initial values for the 
variable $h$ (sampled from a uniform distribution on the  
interval [0,2]) is shown below for illustration. 
 
\begin{figure} 
\centerline{\psfig{figure=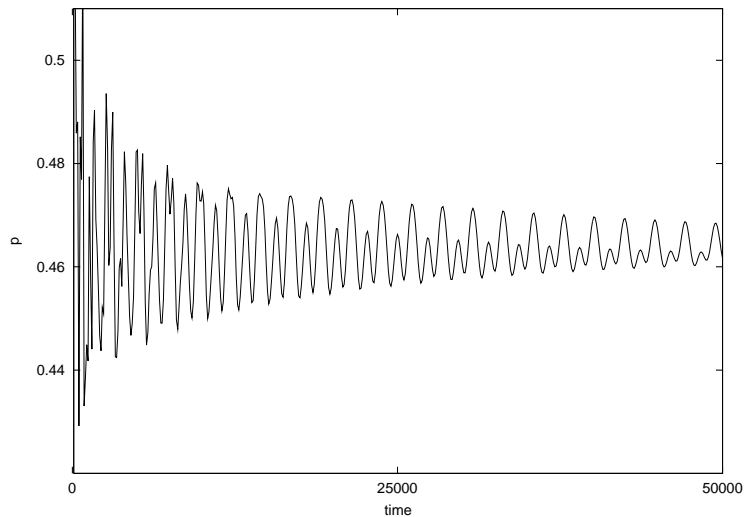,angle=270,height=7cm}}
\vspace{3mm}
\caption{Price variation for single agent. 
	The periodicity is an artifact 
	of the boundary conditions of the lattice.} 
\end{figure} 
  
%%%%%%%%%%%%%%%%%%%%%%%%%%%

\begin{figure} 
\centerline{\psfig{figure=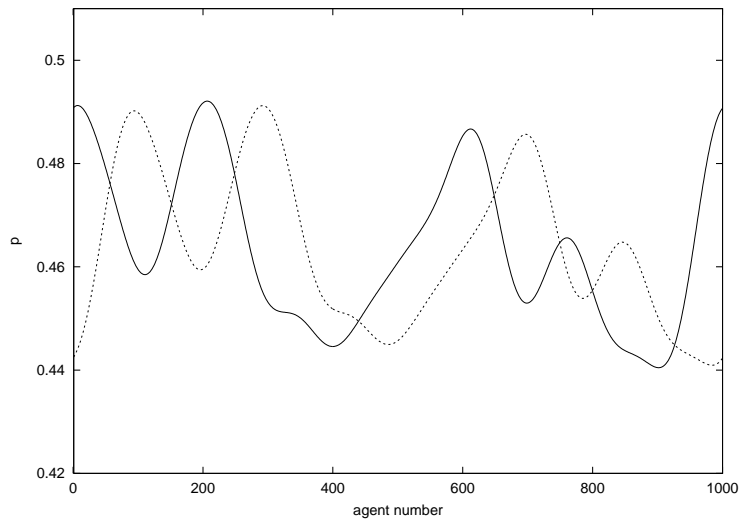,angle=270,height=7cm}}
\vspace{3mm}
\caption{Variation of prices for all agents at two  
	different times, $\tau = 3000$ (solid line) 
	and $\tau=3200$ (dotted line).} 
\end{figure} 
 
Now assume that $h_{n,\tau}$ is a slowly varying function of $(n,\tau)$ 
and that we may think of it as the value of a differentiable function 
$h(x,t)$ in $(x,t)=(n\delta x,\tau \delta t)$. Then, expanding to  
first order in $\delta t$ and second order in $\delta x$,  
we find the diffusion equation 
 
\begin{equation} 
  \frac{\partial h(x,t)}{\partial t}= D \frac{\partial ^2  
  h(x,t)}{\partial x^2} 
  - v \frac{\partial h(x,t)}{\partial x} \enspace ,  
\label{eq:diffus} 
\end{equation} 
where $D = \frac{5}{14}\frac{(\delta x)^2}{\delta t}$ and $v = \frac{3}{7} 
\frac{\delta x}{\delta t}$.  
The generator $T$, of infinitesimal time translations is defined by 
 
\begin{equation} 
  \frac{\partial h(x,t)}{\partial t}= T h(x,t) 
 \enspace . 
\end{equation} 
 
Taking the lattice Fourier  
transformation, the eigenvalues of $T$ are found 
to be $\lambda_k = -k^2 D - i k v$, 
where the periodic boundary condition  
yields $k=\frac{2\pi}{N}\, l$; $l=0,1,\ldots,N-1$. 
The damping time for each mode $k$, 
is given by  $t_k = (k^2 D)^{-1}$. The only mode that is not dampened has 
$k=0$, and is the soft 
``Goldstone mode''~\cite{Hohenberg,Mermin} associated with the  
broken continuous symmetry  with respect to a uniform shift  
of the logarithm of prices in the equilibrium: 
 
All prices can be changed by a common factor, but the amount of 
goods traded will remain the same; as already noted by  
Marx~\cite{Marx} when discussing gold as the unit 
in which prices are expressed, i.e., the measure of value. 
The rest of the modes  
are all damped, and hence the system eventually relaxes to the steady state. 
 
Thus, despite the myopic behavior of the agents, the system manages to  
organize 
itself into the equilibrium. But in contrast to equilibrium theory, we  
obtain 
the temporal relaxation rates towards the equilibrium, as well as  
specific 
absolute values for the individual prices. The value of 
money is fixed by the history of the dynamical process, i.e., 
by the initial condition combined with the actual strategies of 
the bounded rational agents.  
 
Note that the steady states are all Nash equilibria: 
the agents can not change their behavior without loosing utility. 
In the physics analogy, it requires energy for the atoms 
to shift away from the equililibrium, 
as long as their neighbors remain fixed. 
 
Thus, there is a continuous set of perfectly stable Nash equilibria. 
In contrast, the ``search theoretical'' approach~\cite{KW,TW}
yields a single Nash equilibrium, 
with a well defined value of money. The anchoring of 
the value is imposed by requiring the pre-determined number of 
money traders, somewhat artificial and unrealistic, we believe, to spend all  
their money at each transaction.  
If the condition for an agent changes he will change his 
value of money.  
If he, say invents an easier way of production, he will  
lower his price, sell more, and buy more goods, 
the effect being deflation propagating through the system, as 
described by the solution to Eq.~(\ref{eq:diffus}) for a 
delta-function initial condition~\cite{solution}.  
Likewise, if production becomes more difficult for a single agent, or if  
he increases his lust for consumption, this will cause an inflationary wave 
in the the whole system.  
For all agents, except the agent with changed production capability 
and his neighbors, the changes in utility are transient effects. 
In the steady state all these agents will produce and consume 
the same amount of goods as before the change,  
but at a different price level. 
 
How much money is needed to run the economy? Since the agents merely 
seek to preserve a constant amount of money, it does not really matter 
how much money they have, as long as they have sufficient liquidity 
to bridge the interval between buying and selling. One could even imagine 
an economy with no real money at all, as long as the seller would 
provide sufficient credit.  
If, at some point, it were decided by the  
government that dollar bills and coins can not be used for payment, 
this would probably just amount to a nuisance. We would be forced to 
use credit and debit cards and the like, but everything would work just as  
before. One could easily model this situation. The money would merely be 
numbers on accounts.  
 
\section{Conclusion:} 
Here we considered a simple toy model with monopolistic agents.  
In general, economy deals with complicated heterogeneous 
networks of agents, with complicated links to one another, representing 
the particular ``games'' they play with one another. 
We submit 
that the general picture remains the same. At each trade, the agents  
evaluate 
the value of money, by analyzing their particular local situation, 
and act accordingly. The prices charged by the  
agents will 
be constrained by those of the interacting agents. The geometry and topology 
of realistic networks is much more complicated, of course, but the 
continuous symmetry remains. The simple equation Eq.~(\ref{eq:log}) would 
be replaced by a much more complicated avereraging over many interacting 
neighbors. In any case, there is a continuous symmetry related to a global 
change of prices, which is broken by the agents' actions. 
 
It would be interesting 
to study the formation and stability of markets where very many  
distributed players are interested in the same goods, but not generally  
interacting directly with one another.  
Modifications of this network model may also provide a toy laboratory for the  
study of the effects of 
the introduction of the key financial features of credit and bankruptcy  
as well as the control 
problems posed by the governmental role in varying the money supply. 
 
Here the underlying principle for assigning value to money was the dynamical 
breaking of the continuous symmetry, leading to large fluctuations in 
the system when subjected to noise. For stock markets, it is often assumed 
that there is a fundamental value to be discovered by the market. In 
reality, there could easily be so much uncertainty about this value, 
that there would be no real restoring force for price fluctuations 
within a wide range. 
Effectively, we would again have a continuous symmetry, and consequent 
large price fluctuations above and below what could be expected from 
a theory based on rational expectations.


\begin{thebibliography}{99}
\bibitem{solution} The solution to Eq.~(\ref{eq:diffus}) for a  
	delta-function initial condition is:  
	$h(x,t) = \frac{C} 
	{\sqrt{4 \pi D t}}\sum_{j=-\infty}^{\infty} \exp \left(-  
	\frac{(x - jL - vt)^2}{4 D t}\right)$,  
	where $L$ is the system size and $x\in[0,L]$.
\end{thebibliography}
\end{document}